\documentstyle[11pt,epsfig,amsbsy]{article}

\def\be{\begin{equation}}
\def\ee{\end{equation}}
\def\ba{\begin{eqnarray}}
\def\ea{\end{eqnarray}}

\def\ga{\mathrel{\raise.3ex\hbox{$>$\kern-.75em\lower1ex\hbox{$\sim$}}}}
\def\la{\mathrel{\raise.3ex\hbox{$<$\kern-.75em\lower1ex\hbox{$\sim$}}}}

\newcommand{\fr}[2]{\frac{#1}{#2}}

\newcommand{\m}{\rm{m}}

\newcommand{\omde}{\omega_{\rm{de}}}
\newcommand{\Omo}{\Omega_{\rm{m}}^{0}}

\newcommand{\rhom}{\rho_{\rm{m}}}
\newcommand{\rcr}{\rho_{\rm{cr}}}

\newcommand{\vir}{\rm{vir}}
\newcommand{\ta}{\rm{ta}}

\newcommand{\mc}{\rm{cluster}}
\newcommand{\rhoc}{\rho_{\rm{cluster}}}

\newcommand{\AT}{\rm{ArcTan}}
\newcommand{\AS}{\rm{ArcSin}}

\newcommand{\EdS}{\rm{EdS}}
\begin{document}

\baselineskip=16pt
\begin{titlepage}
\begin{center}

\vspace{0.5cm}

\large {\bf Spherical collapse model with and without curvature}
\vspace*{5mm} \normalsize

{\bf Seokcheon Lee$^{\,1,2}$}

\smallskip
\medskip

$^1${\it Institute of Physics, Academia Sinica, \\
Taipei, Taiwan 11529, R.O.C.}

$^2${\it Leung Center for Cosmology and Particle Astrophysics, National Taiwan University, \\ Taipei, Taiwan 10617, R.O.C.}

\smallskip
\end{center}

\vskip0.6in

\centerline{\large\bf Abstract}

We investigate a spherical collapse model with and without the spatial curvature. We obtain the exact solutions of dynamical quantities such as the ratio of the scale factor to its value at the turnaround epoch ($x \equiv \fr{a}{a_{\ta}}$) and the ratio of the overdensity radius to its value at the turnaround time ($y \equiv \fr{R}{R_{\ta}}$) with general cosmological parameters. The exact solutions of the overdensity at the turnaround epoch $\zeta = \fr{\rho_{\mc}}{\rho_{\m}} \Bigl|_{z=z_{\ta}}$ for the different models are also obtained. Thus, we are able to obtain the nonlinear overdensity $\Delta \equiv 1 + \delta_{\rm{NL}} = \zeta \Bigl(\fr{x}{y} \Bigr)^3$ at any epoch for the given model. The nonlinear overdensity at the virial epoch $\Delta_{\vir} = \zeta \Bigl(\fr{x_{\vir}}{y_{\vir}} \Bigr)^3$ is obtained by using the virial theorem and the energy conservation. We obtain that the nonlinear overdensity of the Einstein de Sitter (EdS) Universe $\Delta_{\vir}^{\rm{EdS}}$ is $18 \pi^2 \Bigr(\fr{1}{2\pi} + \fr{3}{4} \Bigr)^2 \simeq 147$ instead of the well known value $18 \pi^2 \simeq 178$. In the open Universe, perturbations are virialized earlier than in flat Universe and thus clusters are denser at the virial epoch. Also the critical density threshold from the linear theory $\delta_{\rm{lin}}^{\rm{EdS}}$ at the virialized epoch is obtained as $\fr{3}{20} (9 \pi + 6)^{\fr{2}{3}} \simeq 1.58$ instead of $\fr{3}{20} (12 \pi)^{\fr{2}{3}} \simeq 1.69$. This value is same for the close and the open Universes. We also show that there is the minimum $z_{\ta}$ value with the given cosmological parameters for each model. We find that the observed quantities at high redshifts are less sensitive between different models. Thus, the high redshift cluster ($z_{\vir} \sim 0.5$) is not a good object to probe the curvature of the Universe. The low redshift cluster ($z_{\vir} \sim 0.07$ {\it i.e.} $z_{\ta} \sim 0.6$) shows the stronger model dependence feature. However, it might be still too small to be distinguished. Because we obtain the analytic forms of dynamical quantities, we are able to estimate the abundances of both virialized and non-virialized clusters at any epoch. Also the temperature and luminosity functions are able to be computed at any epoch. Thus, these analytic forms of $x$, $y$, and $\zeta$ provide the accurate tools for probing the curvature of the Universe. Unfortunately, the current concordance model prefers the almost flat Universe and the above results might be restricted by the academic interests only. However, mathematical structure of the physical evolution equations of the curved space is identical with that of the dark energy with the equation of state $\omde = -\fr{1}{3}$. Thus, we are able to extend these analytic solutions to the general dark energy model and they will provide the useful tools for probing the properties of dark energy.

\vspace*{2mm}

\end{titlepage}


Background evolution equations of physical quantities in a FRW Universe with the matter are given by
\ba && H^2 = \Biggl( \fr{\dot{a}}{a} \Biggr)^2 = \fr{8\pi G}{3} \rhom - \fr{k}{a^2} = \fr{8 \pi G}{3} \rcr \, , \label{H} \\ && \fr{\ddot{a}}{a} = - \fr{4 \pi G}{3} \rhom \, , \label{ddota} \\ && \dot{\rho}_{\rm{m}} + 3 \Bigl( \fr{\dot{a}}{a} \Bigr) \rhom = 0 \, , \label{rhomdot} \ea where $a$ is the scale factor, $\rhom$ is the energy densities of the matter, $\rcr$ is the critical energy density, and $k$ is chosen to be $+1$, $0$, or $-1$ for spaces of constant positive, zero, or negative spatial curvature. In terms of the ratio of the matter density to the critical density $\Omega_{\rm{m}}$, the above Friedmann equation (\ref{H}) becomes \be \fr{k}{H^2a^2} \equiv \Omega_{\rm{k}} = \Omega_{\rm{m}} -1 \, , \label{k} \ee which is valid for all times.

We consider a spherical perturbation in the matter density. $\rhoc$ is the matter density within the spherical overdensity radius $R$. The flatness condition is not held because of the perturbation in the matter (The curvature is not a constant inside the overdensity patch). Thus, we have the another equations governing the dynamics of the spherical perturbation \cite{Peebles}
\ba && \fr{\ddot{R}}{R} = - \fr{4 \pi G}{3} \rhoc \, , \label{ddotR} \\ && \dot{\rho}_{\mc} + 3 \Bigl( \fr{\dot{R}}{R} \Bigr) \rhoc = 0 \, , \label{rhocdot} \ea where $\rho_{\mc}$ is the energy density of the clustering matter. The radius of the overdensity $R$ evolves slower than the scale factor $a$ and reaches its maximum size $R_{\ta}$ at the turnaround epoch $z_{\ta}$ and then the system begins to collapse.

Cosmological parameters and the curvature of the Universe can be constrained from the growth of large scale structure and the abundance of rich clusters of galaxies. There have been numerous works related to this \cite{Gunn,Peebles2,Lahav,Lilje,Lacey,Barrow,Viana,Eke,Kitayama,WS,Horellou}. Most of them reach to the similar conclusions based on the conventional approximate solutions of the background scale factor and of the overdensity radius. It is natural to expect that the correct values for the virial radius and the nonlinear overdensity obtained from the exact solutions might be different from those obtained from the conventional approximate solutions. In order to investigate this, we need to obtain the exact solutions of physical quantities.

Now we adopt the notations in Ref. \cite{WS} to investigate the evolutions of $a$ and $R$ \ba x &=& \fr{a}{a_{\ta}} \, , \label{x} \\ y &=& \fr{R}{R_{\ta}} \, , \label{y} \ea where $a_{\ta}$ is the scale factor of the background evolution at $z_{\ta}$. Then the equations (\ref{H}) and (\ref{ddotR}) are rewritten as \ba \fr{dx}{d \tau} &=& \sqrt{x^{-1} - Q_{\ta}^{-1}} \, , \label{dx} \\ \fr{d^2 y}{d \tau^2} &=& -\fr{1}{2} \zeta y^{-2} \, , \label{ddy} \ea
where $d \tau = H(x_{\ta}) \sqrt{\Omega_{\rm{m}}(x_{\ta})} dt \equiv H_{\ta} \sqrt{\Omega_{\rm{mta}}} dt$, $Q_{\ta} = \fr{\Omega_{\rm{m}}}{\Omega_{\rm{k}}}\Bigl|_{z_{\ta}} \equiv \fr{\Omega_{\rm{mta}}}{\Omega_{\rm{kta}}} = \fr{\Omega_{\rm{mta}}}{\Omega_{\rm{mta}} - 1} = \fr{\Omega_{\rm{m}}^{0}}{\Omega_{\rm{m}}^{0} - 1} (1 + z_{\ta})$, $\zeta = \fr{\rho_{\mc}}{\rho_{\m}} |_{z_{\ta}}$, and $x_{\ta} \equiv x(z_{\ta}) = 1$ from the equation (\ref{x}). $\Omo$ and $\Omega_{\rm{k}}^{0}$ represent the present value of energy density contrasts of the matter and the curvature term, respectively. Equations (\ref{dx}) and (\ref{ddy}) can be solved analytically and we will obtain them.

The analytic solution of Eq. (\ref{dx}) is given by \be \int_{0}^{x} \fr{dx'}{\sqrt{x'^{-1} - Q_{\ta}^{-1}}} = \int_{0}^{\tau} d \tau^{'} \,\, \Rightarrow \,\, \fr{2}{3} x^{\fr{3}{2}} F \Bigl[ \fr{1}{2}\, , \fr{3}{2}\, , \fr{5}{2}\, , \fr{x}{Q_{\ta}} \Bigr] = \tau \, , \label{xtau} \ee where $F$ is the hypergeometric function and we use the boundary condition $x = 0$ when $\tau = 0$ (see Appendix for details). From this equation, the exact turnaround time $\tau_{\ta}$ is given by \be \tau_{\ta} = \fr{2}{3} F \Bigl[ \fr{1}{2}\, , \fr{3}{2}\, , \fr{5}{2}\, , \fr{\Omo -1}{\Omo} (1 + z_{\ta})^{-1} \Bigr] = H_{\ta} \sqrt{\Omega_{\rm{mta}}} t_{\ta} = H_{0} \sqrt{\Omo} (1 + z_{\ta})^{\fr{3}{2}} t_{\ta} \, , \label{tautaw} \ee where we use the fact that $x_{\ta} = 1$, the relations $Q_{\ta} = \fr{\Omega_{\rm{m}}^{0}}{\Omega_{\rm{m}}^{0} - 1} (1 + z_{\ta})$ and $\tau = H_{\ta} \sqrt{\Omega_{\rm{mta}}} t$. This exact analytic form of the turnaround time will be used to investigate the other quantities.
\begin{center}
\begin{figure}
\vspace{1.5cm}
\centerline{
\psfig{file=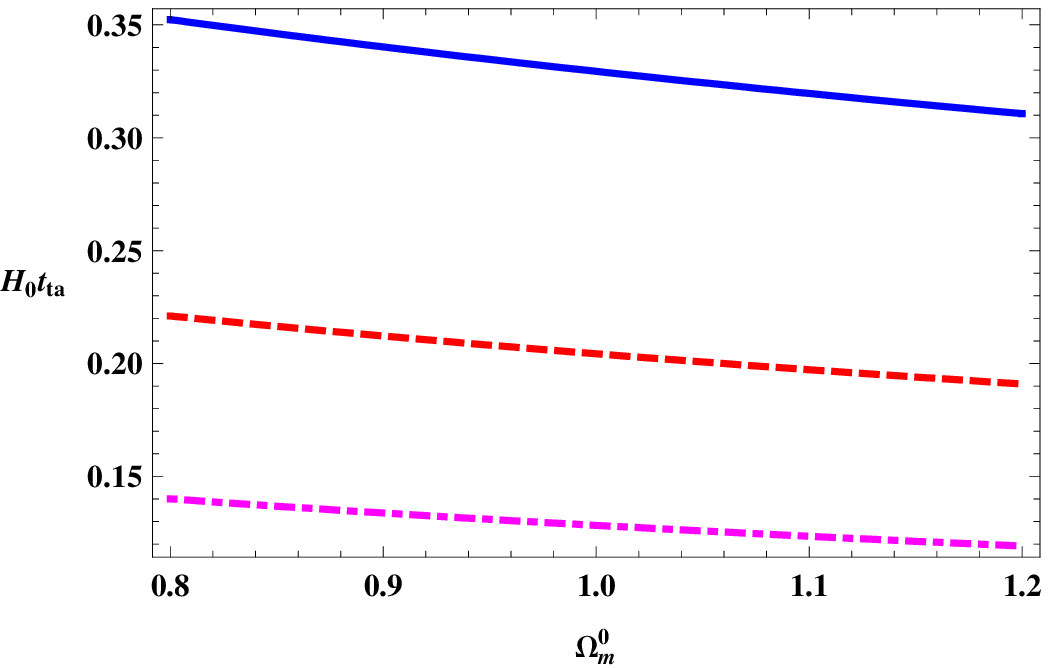, width=6.5cm} \psfig{file=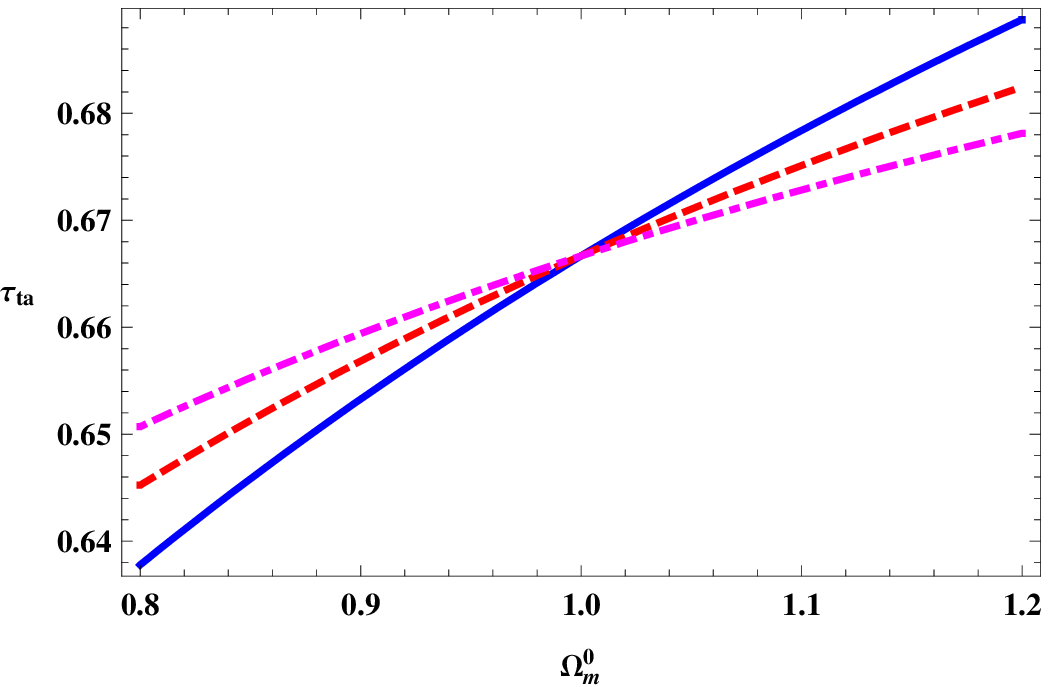, width=6.5cm} }
\vspace{-0.5cm}
\caption{ $t_{\ta}$ and $\tau_{\ta}$ for the different values of $z_{\ta}$. a) $H_{0}t_{\ta}$ verses $\Omo$ for the different values of $z_{\ta} = 0.6\, , 1.2$, and $2.0$ (from top to bottom). b) $\tau_{\ta}$ verses $\Omo$ for the same values of $z_{\ta}$ as in the left panel.} \label{fig1}
\end{figure}
\end{center}

As expected, $\tau_{\ta}$ ($t_{\ta}$) depends on $\Omo$ ({\it i.e.} $\Omega_{\rm{k}}$) and $z_{\ta}$ as given in the Eq. (\ref{tautaw}). We show these properties of $\tau_{\ta}$ ($t_{\ta}$) in Fig. \ref{fig1}. In the left panel of Fig. \ref{fig1}, we show $t_{\ta}$ dependence on the values of $\Omo$ for the different choice of $z_{\ta}$ models. The solid, dashed, and dotdashed lines (from top to bottom) correspond to $z_{\ta} = 0.6, 1.2$, and $2.0$, respectively. Eq. (\ref{xtau}) is the evolution of the background scale factor $a$ and we can interpret it as the age of the Universe is a decreasing function of $\Omo$. Larger $\Omo$ implies faster deceleration, which corresponds a more rapidly expanding universe early on. Also larger $z_{\ta}$ means the earlier formation of the structure and thus gives the smaller $t_{\ta}$. We also show the $\Omo$ dependence of $\tau_{\ta}$ for the values of $z_{\ta}$ in the right panel of Fig. \ref{fig1}. Because $\tau_{\ta} = H_{\ta} \sqrt{\Omega_{\rm{mta}}} t_{\ta}$, $\tau_{\ta}$ becomes larger for the larger values of $\Omo$.

\begin{center}
\begin{figure}
\vspace{1.5cm}
\centerline{
\psfig{file=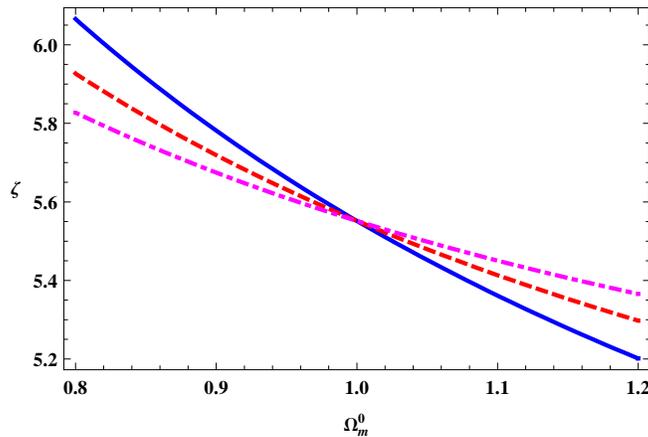, width=8.5cm}}
\vspace{-0.5cm}
\caption{$\zeta$ verses $\Omo$ when we choose $z_{\ta} = 0.6, \, 1.2$, and $2.0$.} \label{fig2}
\end{figure}
\end{center}

The evolution of $y$ given in Eq. (\ref{ddy}) described as the exact analytic solution (see Appendix) \ba && \AS [\sqrt{y} ] - \sqrt{y(1-y)} = \sqrt{\zeta} \tau \,\, ,  \rm{when} \,\, \tau \leq \tau_{\ta} \, , \label{ytaulta} \\ &&
\sqrt{y(1-y)} - \AS [\sqrt{y} ] + \fr{\pi}{2} = \sqrt{\zeta} ( \tau - \tau_{\ta})  \,\, , \rm{when} \,\, \tau \geq \tau_{\ta} \, , \label{ytaugta} \ea where $\tau$ and $\tau_{\ta}$ are given in Eqs. (\ref{xtau}) and (\ref{tautaw}). $\zeta$ can be obtained from this analytic solutions Eq. (\ref{ytaulta}) (or equally from Eq. (\ref{ytaugta})) by using the fact that $y_{\ta} = 1$ \be \zeta = \Biggl( \fr{\pi}{2 \tau_{\ta}} \Biggr)^2 = \Biggl( \fr{3 \pi}{4} \Biggr)^2 \Biggl(F \Bigl[\fr{1}{2}, \fr{3}{2}, \fr{5}{2}, \fr{(\Omo - 1)}{\Omo} (1 + z_{\ta}) ^{-1} \Bigr] \Biggr)^{-2} \, , \label{zeta} \ee where we use Eq. (\ref{tautaw}). When $\Omo = 1$, $F \Bigl[\fr{1}{2}, \fr{3}{2}, \fr{5}{2}, 0 \Bigr] = 1$ and thus $\zeta = (\fr{3 \pi}{4})^2$. This factor $(\fr{3 \pi}{4})^2$ is the well known value of $\zeta$ for the Einstein de Sitter (EdS) Universe ($\Omega_{\rm{m}} = 1$) \cite{Peebles,Kihara}. The general value of $\zeta$ for open or closed Universe is given by Eq. (\ref{zeta}). We show the behavior of $\zeta$ in Fig. \ref{fig2}.

As shown in the equation (\ref{zeta}), $\zeta$ is inversely proportional to $\tau_{\ta}^2$. Thus, $\zeta$ decreases as $\Omo$ increases. Because $\zeta$ is the ratio of $\rho_{\mc}$ to $\rho_{\m}$ at $z = z_{\ta}$, it means that the smaller overdensity takes longer time to turnaround and collapse. In Fig. \ref{fig2}, the solid, dashed, and dotdashed lines correspond to $z_{\ta} = 0.6, \, 1.2$, and $2.0$, respectively. After we obtain the value of $\zeta$, we are able to obtain the values of $x$ and $y$ at any $\tau$ without any ambiguity because the analytic solutions of $x$, $y$, and $\zeta$ given in Eqs. (\ref{xtau}), (\ref{ytaulta}), (\ref{ytaugta}), and (\ref{zeta}) are exact.
\begin{center}
\begin{figure}
\vspace{1.5cm}
\centerline{
\psfig{file=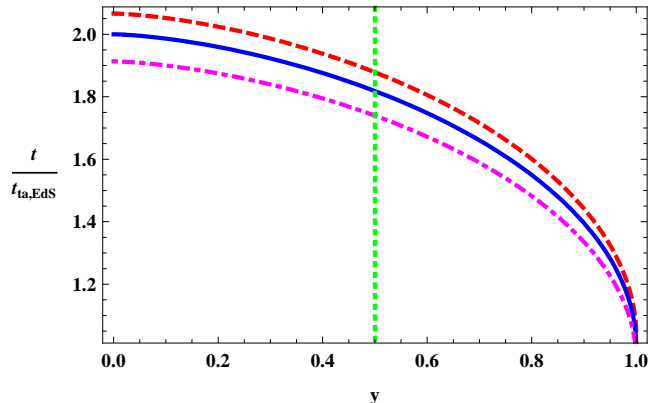, width=8.5cm}}
\vspace{-0.5cm}
\caption{$t$ normalized to the turnaround time for the EdS Universe verses $y$ for close, EdS, and open Universes (from top to bottom).} \label{fig3}
\end{figure}
\end{center}
In order to better understand the above results, it is useful to investigate the virialized times for the different models. From the equation (\ref{ytaugta}), we are able to obtain the collapsing time of each model normalized to the turnaround time for the EdS Universe \be \fr{t}{t_{\ta, \rm{EdS}}} = \fr{2}{\pi} \Bigl( \sqrt{y(1-y)} -\AS [\sqrt{y}] + \pi \Bigr) F \Bigl[\fr{1}{2}\, , \fr{3}{2}\, , \fr{5}{2}\, , Q{\ta}^{-1} \Bigr] \, , \label{ttaEdS} \ee where $t_{\ta, \rm{EdS}}$ is the turnaround time for the EdS Universe. We show this in Fig. \ref{fig3}. Perturbations is virialized earlier in the EdS Universe than in the close Universe, and even earlier in the open Universe. The dashed, solid, and dotdashed lines represent $t$ normalized to the turnaround time of the EdS Universe, $t_{\ta, EdS}$ of close, EdS, and open Universes, respectively. The vertical dotted line depicts the ratio of the virialized radius to the turnaround one $y_{\vir} = \fr{1}{2}$.

If only the matter virializes, then from the virial theorem and the energy conservation we have \be U_{\mc}(z_{\ta}) = \Biggl( U_{\mc} + \fr{R}{2} \fr{\partial U_{\mc}}{\partial R} \Biggr) \Biggl|_{z_{\vir}} = \fr{1}{2} U_{\mc}(z_{\vir}) \, , \label{vir} \ee where $U_{\mc} = - \fr{3}{5} \fr{GM^2}{R}$ is the potential energy associated with the spherical mass overdensity \cite{Peebles,Lahav,Lilje,0505308}. From the above equation we are able to obtain $y_{\vir} = \fr{R_{\vir}}{R_{\ta}}$ for any model \be y_{\vir} = \fr{1}{2} \, . \label{yvir} \ee After replacing $y_{\vir} = \fr{1}{2}$ into Eq. (\ref{ytaugta}), we obtain  \be \tau_{\vir} = \Bigl( \fr{3}{2} + \fr{1}{\pi} \Bigr) \tau_{\ta} = \Bigl( 1 + \fr{2}{3 \pi} \Bigr) F \Bigl[ \fr{1}{2}\, , \fr{3}{2}\, , \fr{5}{2}\, , \fr{1}{Q_{\ta}} \Bigr] \, . \label{tauvir} \ee $x_{\vir}$ is obtained from Eq. (\ref{xtau}) with Eq. (\ref{tauvir}) \be x_{\vir}^{\fr{3}{2}} F \Bigl[ \fr{1}{2}\, , \fr{3}{2}\, , \fr{5}{2}\, , \fr{x_{\vir}}{Q_{\ta}} \Bigr] = \Bigl( \fr{3}{2} + \fr{1}{\pi} \Bigr) F \Bigl[ \fr{1}{2}\, , \fr{3}{2}\, , \fr{5}{2}\, , \fr{1}{Q_{\ta}} \Bigr] \, . \label{xvir} \ee Even though we obtain the analytic expression of $x_{\vir}$ in Eq. (\ref{xvir}), generally this equation can be solved non algebraic way. Thus, it might be useful to have the approximate analytic form of $x_{\vir}$ for the wide ranges of cosmological parameters ($\Omo$ and $z_{\ta}$) if one analyze the form by interpolating $x_{\vir}$ values. We use the exact form in Eq. (\ref{xvir}) though.

Before we move further for the nonlinear overdensity, we investigate the common mistake for $\Delta_{\vir}$ in the EdS Universe. Eqs. (\ref{xtau}), (\ref{ytaugta}), and (\ref{zeta}) can be simplified in EdS Universe because $Q_{\ta}^{-1} = 0$ when $\Omo = 1$. Then the above equations become \ba && \fr{2}{3} x^{\fr{3}{2}} = \tau \, , \label{xtauEdS} \\ && \sqrt{y(1-y)} - \AS [\sqrt{y} ] + \fr{\pi}{2} = \sqrt{\zeta} ( \tau - \fr{2}{3} ) = \fr{\pi}{2} \Bigl(x^{\fr{3}{2}} - 1 \Bigr) \, ,  \label{y133EdS} \\ && \zeta = (\fr{3 \pi}{4})^2 \label{zetaEdS} \, ,\ea
where we use $F[-\fr{1}{2\omde},\fr{1}{2},1-\fr{1}{2\omde},0] = 1$ and $\tau_{\ta} = \fr{2}{3}$. The commonly used assumption to obtain the nonlinear overdensity $\Delta_{c}$ in the EdS Universe is that $\tau_{c} = 2 \tau_{\ta} = \fr{4}{3}$ which is the collapsing time for $y_{c} =0$ even though one uses $y_{\vir} = \fr{1}{2}$ to obtain $\Delta_{c}$. By using this assumption, $x_{c} = 2^{\fr{2}{3}}$ from the equation (\ref{xtauEdS}). Thus, one obtains \be \Delta_{c}^{\EdS} = \zeta \Biggl( \fr{x_{c}}{y_{\vir}} \Biggr)^3 = \Biggl( \fr{3 \pi}{4} \Biggr)^2 \Biggl( \fr{2^{\fr{2}{3}}}{2^{-1}} \Biggr)^{3} = 18 \pi^2 \simeq 178 \, . \label{DeltacEdS1} \ee However, we know the exact relation between $x$ and $y$ and thus we do not need to use the above assumption. If we insert $y_{\vir} = \fr{1}{2}$ in the equation (\ref{y133EdS}), then we obtain the correct $\tau_{\vir} = 1 + \fr{2}{3 \pi} < \tau_{c}$. With this correct value of $\tau_{\vir}$, we obtain the correct $x_{\vir} = (\fr{3}{2} + \fr{1}{\pi})^{\fr{2}{3}}$ by using the Eq. (\ref{xtauEdS}). The correct value of the nonlinear overdensity $\Delta_{\vir}$ for the EdS universe is \be \Delta_{\vir}^{\EdS} = \zeta \Biggl( \fr{x_{\vir}}{y_{\vir}} \Biggr)^3 = \Biggl( \fr{3 \pi}{4} \Biggr)^2 \Biggl( \fr{\Bigl(\fr{3}{2} + \fr{1}{\pi} \Bigr)^{\fr{2}{3}}}{2^{-1}} \Biggr)^{3} = 18 \pi^2 \Biggl( \fr{3}{4} + \fr{1}{2 \pi} \Biggr)^2 \simeq 147 \, . \label{DeltavirEdScor} \ee Thus, the minimum overdensity for the flat Universe is about $147$ instead of $178$. We show this in tables \ref{table1}. As $z_{\ta}$ increases $\Delta_{c}$ approaches to $147$ instead of $178$. Also for the closed universe $\Delta_{\vir}$ can be smaller than this value. From the equation (\ref{x}) we are able to obtain the virial epoch from the given $z_{\ta}$ \be z_{\vir} = \fr{1 + z_{\ta}}{x_{\vir}} -1 \label{zvir} \ee

The ratio of cluster to background density at the virialized epoch $z_{\vir}$ for open and closed Universes becomes \be \Delta_{\vir} = \fr{\rho_{\mc}}{\rho_{\rm{m}}} \Biggl|_{z_{\vir}} = \zeta \Biggl( \fr{x_{\vir}}{y_{\vir}} \Biggr)^3 = 18 \pi^2 \fr{\Bigl(\fr{3}{4} + \fr{1}{2 \pi} \Bigr)^2}{F \Bigl[\fr{1}{2}, \fr{3}{2}, \fr{3}{2}, \fr{x_{\vir}}{Q_{\ta}} \Bigr]^2} \simeq \fr{147}{F \Bigl[\fr{1}{2}, \fr{3}{2}, \fr{3}{2}, \fr{x_{\vir}}{Q_{\ta}} \Bigr]^2} \, , \label{Deltavir} \ee where we use Eqs. (\ref{rhomdot}), (\ref{rhocdot}), and (\ref{zeta}). We also use the fact that $y_{\vir} = \fr{1}{2}$ independent of $Q_{\ta}$. 

\begin{center}
    \begin{table}
    \begin{tabular}{ | c | c | c | c | c | c | c | c | c | c | c | c | c | }
    \hline
      &  \multicolumn{4}{|c|}{$z_{\ta} = 0.6$} & \multicolumn{4}{|c|}{$z_{\ta} = 1.2$} & \multicolumn{4}{|c|}{$z_{\ta} = 2.0$} \\ \cline{2-13}
    $\Omo$ &$\zeta$ & $x_{\vir}$  & $\Delta_{\vir}$ & $z_{\vir}$ &$\zeta$ & $x_{\vir}$  & $\Delta_{\vir}$ & $z_{\vir}$ &$\zeta$ & $x_{\vir}$  & $\Delta_{\vir}$ & $z_{\vir}$  \\ \hline
    $0.80$ & $6.07$ & $1.511$ & $168$ & $0.06$ & $5.93$ & $1.505$ & $162$ & $0.46$
           & $5.83$ & $1.501$ & $158$ & $1.00$ \\ \hline
    $0.95$ & $5.66$ & $1.494$ & $151$ & $0.07$ & $5.63$ & $1.493$ & $150$ & $0.47$
           & $5.61$ & $1.492$ & $149$ & $1.01$ \\ \hline
    $1.00$ & $5.55$ & $1.490$ & $147$ & $0.07$ & $5.55$ & $1.490$ & $147$ & $0.48$
           & $5.55$ & $1.490$ & $147$ & $1.01$ \\ \hline
    $1.05$ & $5.45$ & $1.485$ & $143$ & $0.08$ & $5.48$ & $1.487$ & $144$ & $0.48$
           & $5.50$ & $1.487$ & $145$ & $1.02$ \\ \hline
    $1.20$ & $5.20$ & $1.474$ & $133$ & $0.09$ & $5.30$ & $1.478$ & $137$ & $0.49$
           & $5.37$ & $1.481$ & $140$ & $1.03$ \\ \hline
    \end{tabular}
    \caption{$\zeta$, $x_{\vir}$, $\Delta_{\vir}$, and $z_{\vir}$ with the given values of cosmological parameters for the open, flat, and closed Universes. Perturbations reach turnaround and virialization earlier in the flat Universe than in the close Universe, and even earlier in the open Universe. Thus, clusters are denser at virial epoch in flat Universe than in the close Universe and even denser in the open Universe.}
    \label{table1}
    \end{table}
\end{center}

From the equation (\ref{Deltavir}), we are able to investigate the linear perturbation at early epoch $\Delta \,\, \stackrel{\tau\rightarrow 0}{\longrightarrow} \,\, 1 + \delta_{\rm{lin}}$ (see Appendix for details). \be \Delta \equiv 1 + \delta_{\rm{NL}} \stackrel{\tau\rightarrow 0}{\longrightarrow} \, 1 + \delta_{\rm{lin}} = 1 + \fr{3}{5} \Bigl(\sqrt{\zeta} \fr{3}{2} \tau \Bigr)^{\fr{2}{3}} = 1 + \fr{3}{5} \Bigl(\sqrt{\zeta} \sqrt{\Omega_{\rm{mta}}} \fr{t}{t_{\ta}} \Bigr)^{\fr{2}{3}} \, . \label{Deltadelta} \ee It is equal to the famous result for EdS Universe ($\sqrt{\zeta} = \fr{3 \pi}{4}$ and $\Omega_{\rm{mta}} = 1$), $\delta_{\rm{lin}} = \fr{3}{5} \Bigr( \fr{3 \pi}{4} \fr{t}{t_{\ta}} \Bigr)^{\fr{2}{3}}$ \cite{Peebles}. Again, there is a common mistake for the value of the critical density threshold $\delta_{\rm{lin}}(z_{\vir})$. In EdS Universe, $\tau_{c} = \fr{4}{3}$ and one obtains $\delta_{\rm{lin}}(z_{c})$ from Eq. (\ref{Deltadelta}) \be \delta_{\rm{lin}}(z_{c}) = \fr{3}{5} \Bigl( \fr{3 \pi}{4} \Bigr)^{\fr{2}{3}} \Bigl( \fr{3}{2} \times \fr{4}{3} \Bigr)^{\fr{2}{3}} = \fr{3}{20} (12 \pi)^{\fr{2}{3}} \simeq 1.69 \, . \label{deltac} \ee However, we show that the correct virialized epoch gives $\tau_{\vir} = \Bigl(\fr{3}{2} + \fr{1}{\pi} \Bigr) \tau_{\ta}$ from Eq. (\ref{tauvir}). Thus, the correct $\delta_{\rm{lin}}(z_{\vir})$ value is \be \delta_{\rm{lin}}(z_{\vir}) = \fr{3}{5} ( \sqrt{\zeta})^{\fr{2}{3}} \Biggl( \Bigl( \fr{3}{4} + \fr{9 \pi}{8} \Bigr) \fr{1}{\sqrt{\zeta}} \Biggr)^{\fr{2}{3}} = \fr{3}{20} (6 + 9 \pi)^{\fr{2}{3}} \simeq 1.58 \, , \label{deltavir} \ee where we use $\sqrt{\zeta} = \fr{\pi}{2 \tau_{\ta}}$. The above result given in Eq. (\ref{deltavir}) is true for with and without curvature and thus it is valid for the close, flat, and open Universes. After we obtain the critical density threshold $\delta_{\rm{lin}}(z_{\vir})$ we are able to obtain $\delta_{\rm{lin}}$ at any epoch by using the relation \be \delta_{\rm{lin}}(z) = \fr{D_{g}(z)}{D_{g}(z_{\vir})} \delta_{\rm{lin}}(z_{\vir}) \label{deltaz} \ee where $D_{g}$ is the linear growth factor. There is the exact analytic form of $D_{g}$ for the dark energy with the equation of state $\omde = -\fr{1}{3}$ \cite{SK1}. Mathematically, this form of $D_{g}$ is identical with that of curved space and thus we can adopt this form of $D_{g}$ in these models.

From the analytic forms of dynamical quantities $x$, $y$, and $\zeta$, it is straightforward to  estimate the abundances of both virialized and non-virialized clusters at any epoch. Also the temperature and luminosity functions are able to be computed at any epoch \cite{SK2}. Thus, these analytic forms provide an accurate tool for probing the effect of the curvature on the clustering. As we mentioned, the mathematical structure of the physical evolution equations of the curved space is identical with that of the dark energy with the equation of state $\omde = -\fr{1}{3}$. Thus, these analytic solutions give the guideline for the extension of them to those of the general dark energy model and they will provide the useful tools for probing the properties of dark energy \cite{SK3}.

\appendix
\section{Appendix}
\setcounter{equation}{0}
First, we derive the exact solution of $\tau$ as a function of $x$ given in Eq. (\ref{xtau}). After we replace the variables $Z= \fr{x}{Q_{\ta}}$ and $T = \fr{x'}{x}$, the integral in Eq. (\ref{xtau}) becomes \cite{Abramowitz} \be \int_{0}^{x} \fr{dx'}{\sqrt{x'^{-1} - Q_{\ta}^{-1}}} = x^{\fr{3}{2}} \int_{0}^{1} T^{\fr{1}{2}} (1 - ZT)^{-\fr{1}{2}} dT = \fr{2}{3} x^{\fr{3}{2}} F \Bigl[\fr{1}{2}, \fr{3}{2}, \fr{5}{2}, \fr{x}{Q_{\ta}} \Bigr] \, , \label{xtaut} \ee where $F$ is the hypergeometric function and we use the gamma function relation $\Gamma[1+b] = b \Gamma[b]$.

We also derive the exact analytic solution of $\tau$ as a function of $y$ given in Eq. (\ref{ddy}). This equation is solved as \cite{HB} \be \fr{d y^2}{d \tau^2} =-\fr{1}{2} \zeta y^{-2} \equiv f(y) \,\, \Rightarrow \,\, \int_{0}^{y} \fr{dy'}{\sqrt{-c_{1} + 2 \int^{y'} f(y'') dy''}} = c_{2} \pm \tau \, . \label{y13} \ee We use $2 \int^{y'} f(y'') dy'' = \zeta y'^{-1}$ to obtain \be \int_{0}^{y} \fr{dy'}{\sqrt{-c_{1} + \zeta y'^{-1}}} = c_{2} \pm \tau \, . \label{y132} \ee After replacing $Z = \fr{c_{1}}{\zeta} y$ and $T = \fr{y'}{y}$, the LHS of Eq. (\ref{y132}) becomes \be \int_{0}^{y} \sqrt{\fr{y'}{\zeta (1 -y')}} dy' = \fr{y^{\fr{3}{2}}}{\sqrt{\zeta}} \int_{0}^{1} T^{\fr{1}{2}} (1 - \fr{c_{1}}{\zeta} y T)^{-\fr{1}{2}} dT = \fr{2}{3} \fr{1}{\sqrt{\zeta}} y^{\fr{3}{2}} F \Bigl[\fr{1}{2}, \fr{3}{2}, \fr{5}{2}, \fr{c_{1}}{\zeta} y \Bigr] \, . \label{y13t} \ee We use the relation $ \fr{c_{1}}{\zeta} y F [\fr{1}{2}, \fr{3}{2}, \fr{5}{2}, \fr{c_{1}}{\zeta} y] = \fr{3}{2} F [\fr{1}{2}, \fr{1}{2}, \fr{3}{2}, \fr{c_{1}}{\zeta} y] - \fr{3}{2} (1 - \fr{c_{1}}{\zeta} y) F [\fr{1}{2}, \fr{3}{2}, \fr{3}{2}, \fr{c_{1}}{\zeta} y]$, $F[\fr{1}{2}, \fr{1}{2}, \fr{3}{2}, y^2] = y^{-1} \rm{ArcSin}[y]$, and $F[a,b,b,y] = (1-y)^{-a}$ to obtain \ba
&& \fr{\sqrt{\zeta y}}{c_1} \Biggl( \sqrt{\fr{\zeta}{c_1 y}} \rm{ArcSin} \Bigl[\sqrt{\fr{c_1 y}{\zeta}} \Bigr] - \sqrt{1 - \fr{c_1 y}{\zeta}} \Biggr) = c_{2l} + \tau \,\, \rm{when} \,\, \tau \leq \tau_{\ta} \, , \label{ylta} \\
&& \fr{\sqrt{\zeta y}}{c_1} \Biggl( \sqrt{\fr{\zeta}{c_1 y}} \rm{ArcSin} \Bigl[\sqrt{\fr{c_1 y}{\zeta}} \Bigr] - \sqrt{1 - \fr{c_1 y}{\zeta}} \Biggr) = c_{2g} - \tau \,\, \rm{when} \,\, \tau \geq \tau_{\ta} \, , \label{ygta} \ea where $\AS$ represents arcsine function.
We choose the positive (negative) sign in $\tau$ to be consistent with the fact that $R$ ({\it i.e.} $y$) increases (decreases) as $\tau$ increases before (after) it reaches to the maximum radius $R_{\ta}$ ( $y_{\ta} =1$) at $\tau_{\ta}$. Integral constants $c_{1}$, $c_{2l}$, and $c_{2g}$ are obtained from the boundary conditions $\fr{d y}{d \tau} |_{\ta} = 0$, $y(\tau=0) = 0$, and $y_{\ta} = 1$ \ba \fr{d y}{d \tau} \Biggl|_{\ta} &=& \pm \sqrt{-c_{1} + \fr{\zeta}{y_{\ta}}} = 0 \,\, \Rightarrow \,\, c_{1} = \zeta \, , \label{dydtau} \\ 0 &=& c_{2l} \label{ddytaul} \\ \fr{\pi}{2 \sqrt{\zeta}} &=& c_{2g} - \tau_{\ta} \, , \label{ddytaug} \ea where we use the fact that $\AS [1] = \fr{\pi}{2}$. $\tau_{\ta}$ is obtained from Eq. (\ref{tautaw}) \be \tau_{\ta} = \fr{2}{3} F \Bigl[ \fr{3}{2}, \fr{1}{2}, \fr{5}{2}, - \fr{1}{Q_{\ta}} \Bigr] \, . \label{tauta} \ee From Eqs. (\ref{ddytaul}), (\ref{ddytaug}), and (\ref{tauta}), we obtain the exact analytic solution of $y(\tau)$ \ba && \AS [\sqrt{y} ] - \sqrt{y(1-y)} = \sqrt{\zeta} \tau \,\, , \rm{when} \,\, \tau \leq \tau_{\ta} \, . \label{y133appl} \\&& \sqrt{y(1-y)} - \AS [\sqrt{y} ] + \fr{\pi}{2} = \sqrt{\zeta} ( \tau - \tau_{\ta} ) \,\, , \rm{when} \,\, \tau \geq \tau_{\ta} \, . \label{y133appg} \ea
If we use the relationships $ -\rm{ArcSin} \Bigl[\sqrt{y} \Bigr] + \fr{\pi}{4} = \fr{1}{2} \AT \Bigl[\fr{1-2y}{2} \sqrt{\fr{1}{y(1-y)}} \Bigr] = -\fr{2}{3} y^{\fr{3}{2}} F \Bigl[ \fr{1}{2}, \fr{3}{2}, \fr{5}{2}, y \Bigr] - \sqrt{y(1-y)} + \fr{\pi}{4}$, then we are able to rewrite the above equations (\ref{y133appl}) and (\ref{y133appg}) as \ba && - \sqrt{y(1-y)} - \fr{1}{2} \AT \Bigl[\fr{1-2y}{2} \sqrt{\fr{1}{y(1-y)}} \Bigr] + \fr{\pi}{4} \nonumber \\ &&
= \fr{2}{3} y^{\fr{3}{2}} F \Bigl[ \fr{1}{2}, \fr{3}{2}, \fr{5}{2}, y \Bigr] = \sqrt{\zeta} \tau  \,\, , \rm{when} \,\, \tau \leq \tau_{\ta} \, , \label{y133app2l} \\ && \sqrt{y(1-y)} + \fr{1}{2} \AT \Bigl[\fr{1-2y}{2} \sqrt{\fr{1}{y(1-y)}} \Bigr] + \fr{\pi}{4} \nonumber \\ && = \fr{\pi}{2} - \fr{2}{3} y^{\fr{3}{2}} F \Bigl[ \fr{1}{2}, \fr{3}{2}, \fr{5}{2}, y \Bigr] = \sqrt{\zeta} ( \tau - \tau_{\ta} ) \,\, , \rm{when} \,\, \tau \geq \tau_{\ta} \, . \label{y133app2g} \ea Thus, all of these three representations are equal to each other.

We also check the perturbation is linear at early epoch $\Delta \,\, \stackrel{\tau\rightarrow 0}{\longrightarrow} \,\, \delta_{\rm{lin}}$. We use equations (\ref{xtaut}) and (\ref{y133appl}) because $\tau \leq \tau_{\ta}$ in this case. Thus when we choose the limit $\tau \rightarrow 0$, we obtain $\fr{2}{3} x^{\fr{3}{2}} \simeq \tau$ and $\fr{2}{3} y^{\fr{3}{2}} + \fr{1}{5} y^{\fr{5}{2}} \simeq \sqrt{\zeta} \tau$. Thus we obtain \be \Delta \equiv 1 + \delta_{\rm{NL}} = \zeta \Bigl( \fr{x}{y} \Bigr)^3 \, \stackrel{\tau\rightarrow 0}{\longrightarrow} \, \zeta \Biggl( \fr{1}{\sqrt{\zeta}} \Bigl( 1 + \fr{3}{10} y \Bigr) \Biggr)^2 \simeq 1 + \fr{3}{5} y \,\, \Rightarrow \,\, \delta_{\rm{NL}} \rightarrow \delta_{\rm{lin}} = \fr{3}{5} (\sqrt{\zeta})^{\fr{2}{3}} \Bigl(\fr{3}{2} \tau \Bigr)^{\fr{2}{3}} \label{Deltadeltaapp} \ee If we use $\tau = H_{\ta} \sqrt{\Omega_{\rm{mta}}} t$ and $H_{\ta} = \fr{2}{3} t_{\ta}^{-1}$ in the matter dominated epoch, then we obtain $\delta_{\rm{lin}} = \fr{3}{5} \Bigl( \sqrt{\zeta} \sqrt{ \Omega_{\rm{mta}} } \fr{t}{t_{\ta}} \Bigr)^{\fr{2}{3}}$. In EdS Universe ($\sqrt{\zeta} = \fr{3 \pi}{4}$ and $\Omega_{\rm{mta}} = 1$), this gives the famous result $\delta_{\rm{lin}} = \fr{3}{5} \Bigl( \fr{3 \pi}{4} \fr{t}{t_{\ta}} \Bigr)^{\fr{2}{3}}$. For the close and open Universe, $\delta_{\rm{lin}}(z_{\vir}) = \fr{3}{5} \Bigl( \sqrt{\zeta} \fr{3}{2} \tau_{\vir} \Bigr)^{\fr{2}{3}} = \fr{3}{20} ( 6 + 9 \pi )^{\fr{2}{3}}$ which is equal to that of the EdS one.

We thanks K. Umetsu for useful comment.

\end{document}